\documentclass[prb,aps,twocolumn,footinbib,floatfix,10pt,longbibliography,superscriptaddress]{revtex4-1}

\usepackage{comment}
\usepackage{chemformula} 
\usepackage[T1]{fontenc} 
\usepackage{graphicx}
\usepackage{dcolumn}
\usepackage{lipsum}
\usepackage{tabu} 
\usepackage{mathtools}

\usepackage{braket}
\usepackage{color}
\usepackage[11pt]{moresize}
\usepackage{anyfontsize}
\usepackage{bbding}
\usepackage{amsmath}
\usepackage{xcolor}
\usepackage{braket}
\usepackage[normalem]{ulem}

\begin{document} 

\title{On-chip indistinguishable photons using III-V nanowire/SiN hybrid integration}

\author{Edith Yeung}
\affiliation{National Research Council Canada, Ottawa, Ontario, Canada, K1A 0R6.}
\affiliation{University of Ottawa, Ottawa, Ontario, Canada, K1N 6N5.}
\author{David B. Northeast}
\affiliation{National Research Council Canada, Ottawa, Ontario, Canada, K1A 0R6.}
\author{Jeongwan Jin}
\affiliation{National Research Council Canada, Ottawa, Ontario, Canada, K1A 0R6.}
\author{Patrick Laferri{\`e}re}
\affiliation{National Research Council Canada, Ottawa, Ontario, Canada, K1A 0R6.}
\author{Marek Korkusinski}
\affiliation{National Research Council Canada, Ottawa, Ontario, Canada, K1A 0R6.}
\affiliation{University of Ottawa, Ottawa, Ontario, Canada, K1N 6N5.}
\author{Philip J. Poole}
\affiliation{National Research Council Canada, Ottawa, Ontario, Canada, K1A 0R6.}
\author{Robin L. Williams}
\affiliation{National Research Council Canada, Ottawa, Ontario, Canada, K1A 0R6.}
\author{Dan Dalacu}
\affiliation{National Research Council Canada, Ottawa, Ontario, Canada, K1A 0R6.}
\affiliation{University of Ottawa, Ottawa, Ontario, Canada, K1N 6N5.}

\begin{abstract}

We demonstrate on-chip generation of indistinguishable photons based on a nanowire quantum dot. From a growth substrate containing arrays of positioned-controlled single dot nanowires, we select a single nanowire which is placed on a SiN waveguide fabricated on a Si-based chip. Coupling of the quantum dot emission to the SiN waveguide is via the evanescent mode in the tapered nanowire. Post-selected two-photon interference visibilities using continuous wave excitation above-band and into a p-shell of the dot were 100\%, consistent with a single photon source having negligible multi-photon emission probability. Visibilities over the entire photon wavepacket, measured using pulsed excitation, were reduced by a factor of 5 when exciting quasi-resonantly and by a factor of 10 for above-band excitation. The role of excitation timing jitter, spectral diffusion and pure dephasing in limiting visibilities over the temporal extent of the photon is investigated using additional measurements of the coherence and linewidth of the emitted photons.

\end{abstract}

\maketitle 

\section{Introduction}

High interference visibility between two single photons incident on separate input ports of a 50/50 beamsplitter, the Hong-Ou-Mandel (HOM) effect\cite{HOM}, establishes the indistinguishable nature of the photons, an essential requirement in most photonic quantum technologies\cite{OBrien_NP2009}. Epitaxial semiconductor quantum dots offer a solid-state solution for deterministically generating indistinguishable photons \cite{Santori_NAT2002} with state-of-the-art sources demonstrating two-photon interference (TPI) visibilities in excess of 95\% \cite{He_NN2013,Ding_PRL2016,Somaschi_NP2016} for pulse separations of over $14\mu$s \cite{Wang_PRL2016} in devices with efficiencies of up to 57\% \cite{Tomm_NN2021}. These sources were designed to emit out-of-plane, whereas a key advantage of solid-state emitters is the ability to integrate them with on-chip photonic circuitry\cite{Hepp_AQT2019}. An integrated platform, whereby multiple sources generate indistinguishable photons \cite{Patel_NP2010,Kim_NL2016b} propagating within on-chip photonic circuitry \cite{Ellis_APL2018}, is a long-term challenge addressing scalability requirements of complex quantum processing schemes \cite{OBrien_NP2009}.

Two distinct technologies for generating on-chip indistinguishable photons using quantum dots are currently being pursued.  In monolithic approaches \cite{Dietrich_LPR2016}, the quantum dot is embedded within photonic crystal waveguides \cite{Kalliakos_APL2014,Kalliakos_APL2016} or suspended nanobeams \cite{Prtijaga_APL2016} fabricated from the same III-V material system. Hybrid platforms \cite{Elshaari_NP2020}, on the other hand, combine III-V-based quantum dot systems with Si-based integrated circuits in which the dot emission is coupled to waveguide structures using either direct butt-coupling \cite{Ellis_APL2018} or evanescent fields \cite{Schnauber_NL2018}. Initial experiments \cite{Kalliakos_APL2014,Kalliakos_APL2016,Prtijaga_APL2016,Schnauber_NL2019} demonstrating on-chip generation of indistinguishable photons using the above approaches relied on post-selected TPI visibilities where excitation was provided by a continuous wave (CW) laser. More recently, non-post-selected measurements made using resonant pulsed excitation have shown TPI visibilities of $V > 90$\% between sequentially emitted photon separated by a few nanoseconds \cite{Kirsanske_PRB2017,Dusanowski_PRL2019,Uppu_NC2020,Dusanowski_NL2020} up to almost 800\,ns \cite{Uppu_SA2020}.

In this work we report on a deterministic hybrid technique for generating in-plane indistinguishable photons. We use position-controlled nanowire quantum dots \cite{Laferriere_SR2022} incorporated in photonic nanowire waveguides that are designed for efficient evanescent coupling to SiN waveguides\cite{Mnaymneh_AQT2020}. The approach is similar to previous techniques employing evanescent coupling of III-V structures to underlying Si-based photonic structures \cite{Kim_NL2017,Davanco_NC2017,Schnauber_NL2018}, except here the adiabatic mode transfer is not realized through geometries defined by lithography but rather by a taper introduced in the photonic nanowire during growth \cite{Dalacu_NANOM2021}. The hybrid construction relies on a pick and place technique whereby individual nanowires are picked up from the III-V growth substrate and placed on a SiN waveguide fabricated on a Si wafer. The transfer is carried out in a scanning electron microscope (SEM) which provides a placement precision of a few nanometers; a positioning control sufficient to achieve optimal nanowire-waveguide coupling.

\begin{figure}
    \includegraphics[width=8.5cm,clip=true]{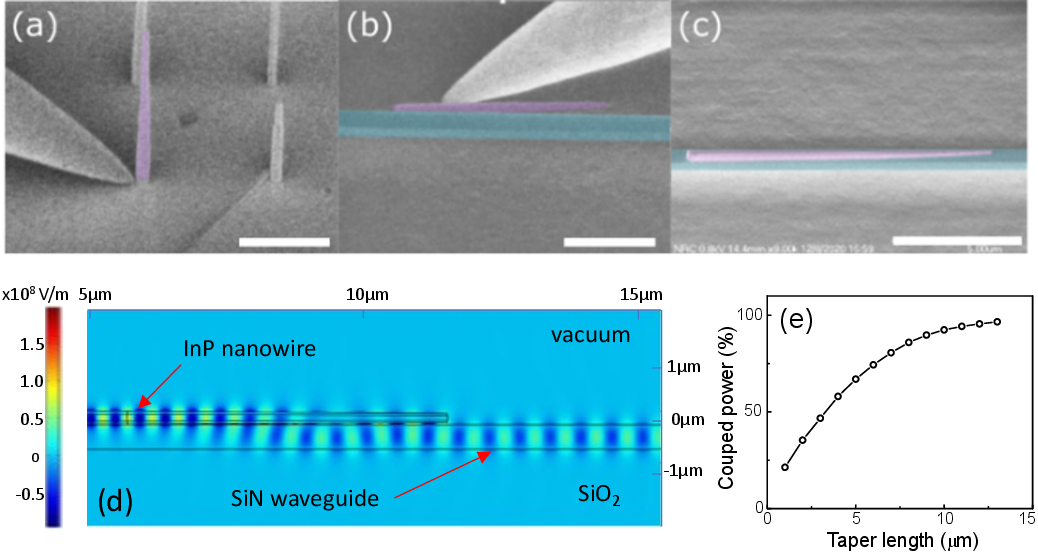}
    \caption{Pick and place technique: (a) Nanowire (purple) picked up from growth substrate using a nanomanipulator probe. (b) Nanowire placed next to a SiN waveguide (blue). (c) Integrated InP nanowire-SiN ridge waveguide device. Scale bars are $5\mu$m. (d)  Simulation of the electric field $E_x$ of the fundamental waveguide mode propagating though a cross-section of the hybrid device. (e) Calculated nanowire to waveguide mode coupling as a function of the nanowire taper length.}
    \label{Figure1}
\end{figure}

We obtain peak TPI visibilities of $V=19.2\%$ over the temporal extent of the emitted photons when exciting non-resonantly. To gain insight into the the different mechanisms limiting the observation of high visibilities, we perform additional experiments sensitive to decoherence processes in two-level systems: first-order correlation, $g^{(1)}(\tau)$, and high resolution spectroscopic measurements. 


\section{Chip-integrated nanowire source}

The quantum dots used in this work are sections of InAsP incorporated within position-controlled InP nanowires grown using gold-catalyzed vapour-liquid-solid epitaxy described in detail in Refs.~\citenum{Dalacu_NT2009,Dalacu_APL2011,Dalacu_NL2012}. The nanowires are clad with an InP shell to create a photonic nanowire with a base diameter of 250\,nm that supports single mode waveguiding of 1.3\,eV photons. The photonic nanowire is tapered to a tip diameter of 100\,nm over a $\sim 15\,\mu$m length (see Fig.~\ref{Figure1}(a)) to enable adiabatic mode transfer, discussed below. The low-loss photonic circuitry was fabricated in SiN grown by low-pressure chemical vapour deposition with measured propagation loss $\sim~0.48$\,dB/cm at 965\,nm. Waveguide dimensions were 400\,nm wide and 485\,nm thick with SiO$_2$ below and above, designed to support a single polarization mode at 1.3\,eV. The SiN waveguides were terminated at etched facets of the Si chip where the width was tapered to 250\,nm for efficient coupling to single mode lensed fibres (SMLF) with measured coupling losses of $\sim2.15$\,dB/facet.

\begin{figure}
    \includegraphics[width=8.5cm,clip=true]{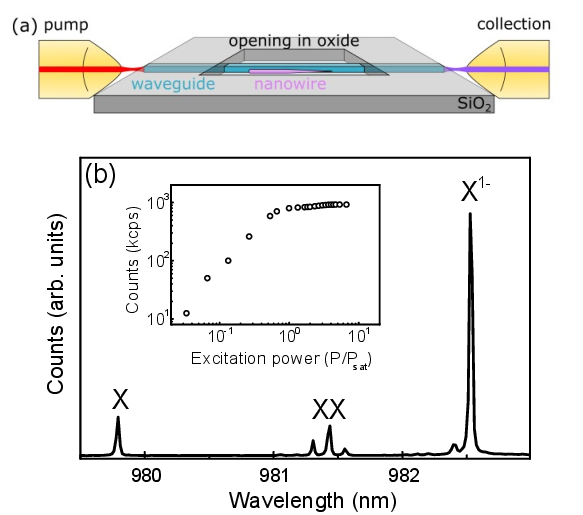}
    \caption{(a) Schematic of the pump and collection scheme. (b) Photoluminescence spectrum from a hybrid integrated device. The inset shows the detected count rates as a function of pump power using pulsed quasi-resonant excitation at 80\,MHz.}
    \label{Figure2}
\end{figure}

The nanowires were transferred to the Si chip pre-fabricated with the SiN photonic circuitry using an SEM-based nanomanipulator as shown in Fig.~\ref{Figure1}(a)-(c). A single nanowire is picked up from the growth substrate with a tungsten tip controlled by piezo-motors and then moved to the Si chip mounted next to the growth substrate. The nanowire is then placed either on or beside a selected SiN waveguide which has been exposed by opening a $50\times 50\,\mu$m$^2$ window in the top oxide, see Fig.~\ref{Figure2}(a). As mentioned above, the nanowires are tapered to promote adiabatic mode transfer from the InP nanowire to the SiN waveguide. In Fig.~\ref{Figure1}(d) we show a simulation of the HE$_{11}$ mode in the nanowire transfer to the TE mode in the SiN waveguide. From this we calculate a transfer efficiency in excess of 90\% for the nanowire geometry specified above but with a waveguide design that supports both TE and TM polarizations.

The chip was cooled to 4\,K in a fibre-coupled closed-cycle He cryostat equipped with xyz piezo stages for fibre-waveguide alignment. The nanowires were optically excited through the waveguide via a SMLF. The emission was collected on the other end of the waveguide through another SMLF and directed to a fibre-coupled grating spectrometer equipped with a nitrogen-cooled charge-coupled device for spectrally resolved measurements or to superconducting nanowire single photon detectors (SNSPD, timing jitter 100\,ps) via a fibre-coupled tunable filter (bandwidth = 0.1\,nm) for measurements on single lines. An s-shell photoluminescence (PL) spectrum from the on-chip source is shown in Fig.~\ref{Figure2}(b) and displays the typical exciton complexes ($X$, $XX$, $X^{1-}$) observed from such nanowire quantum dots\cite{Laferriere_APL2021}.

In this article, we focus on the $X^{1-}$ emission. We first determine the efficiency of single photon generation from the device. For pulsed above-band excitation at 80\,MHz, we measure a maximum of 0.919\,Mcps at the SNSPD detector at an excitation power $P=P_\mathrm{sat}$ that saturates the transition, see inset in Fig.~\ref{Figure2}(b). Taking into account the optical throughput of the system (8.1\%) and the detector efficiency (88.5\%), both measured at 980\,nm, we obtained a first-lens count rate of 12.9\,Mcps, corresponding to source efficiency of $\eta_s\sim16.1\%$. To estimate the dot to SiN waveguide coupling, we consider ($i$) a calculated dot-HE$_{11}$ coupling of $\beta =95$\%\cite{Dalacu_NT2019}, ($ii$) a 50\% loss of photons for emission directed towards the base of the nanowire, ($iii$) a 50\% loss due to the waveguide design which only supports one of the two polarization modes in the nanowire, and ($iv$) a 20\% loss of photons emitted into phonon sidebands\cite{Dalacu_NT2019} which are filtered-out. Taking into account these losses, we obtain an evanescent coupling efficiency of $\eta_c\sim84.8\%$, slightly lower than both the calculations shown in Fig.~\ref{Figure1}(e) and our best measured results to date, $\eta_c\sim93\%$ (see Ref.~\citenum{Yeung_2021}). The lower than predicted values may be associated with emission into other charge complexes\cite{Laferriere_SR2022}, e.g. the neutral complexes $X$ and $XX$, evident in Fig.~\ref{Figure2}(b).

\section{TPI measurements: CW excitation}

\begin{figure}
    \includegraphics[width=9cm,clip=true]{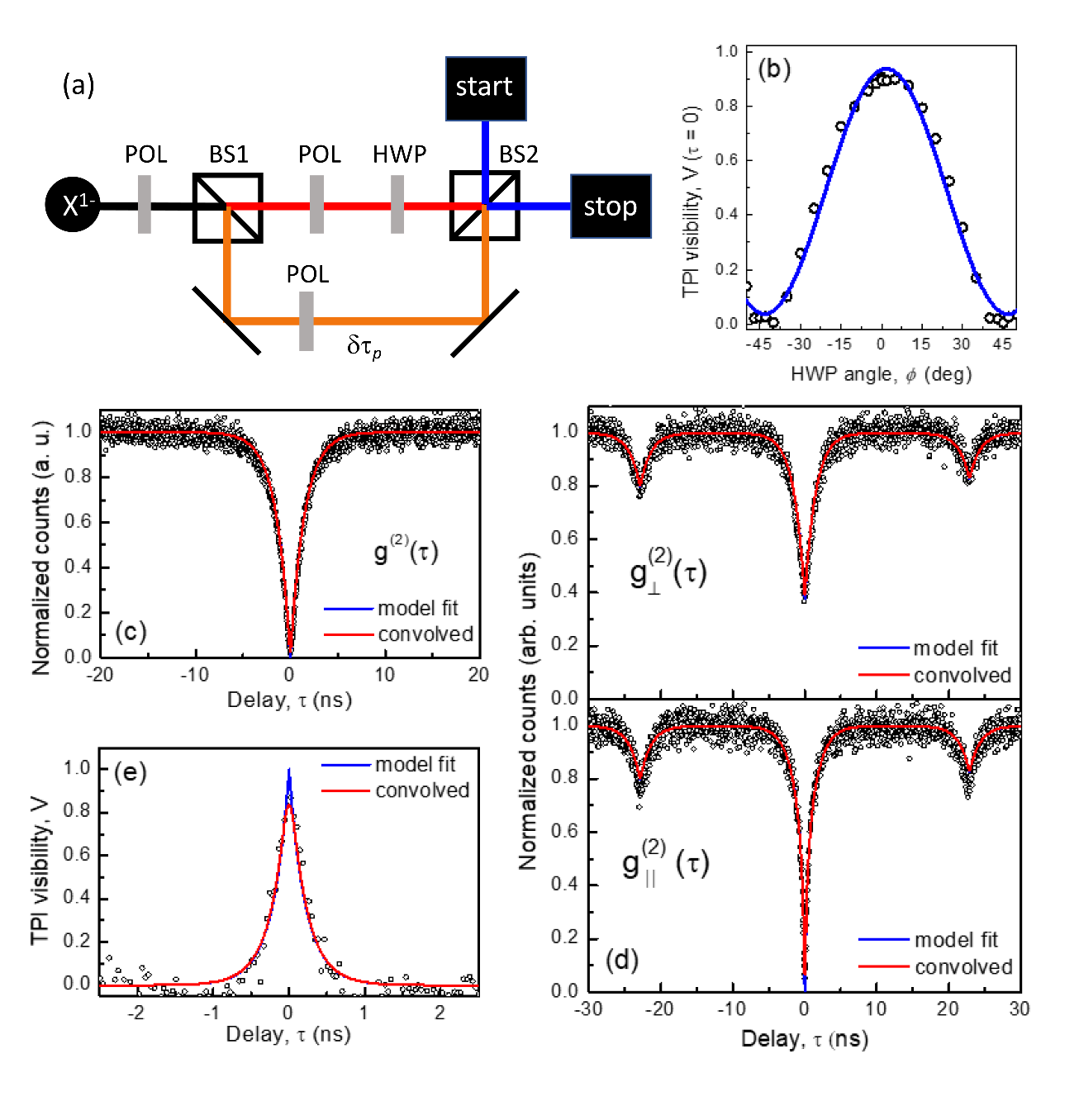}
    \caption{(a) Schematic of Hong-Ou-Mandel (HOM) setup. (b) Raw TPI visibility at $\tau=0$\,ns as a function of HWP angle $\phi$ measured using above-band CW excitation at $P_\mathrm{sat}$ and a $\delta\tau_p=1.84$\,ns delay in one arm of the MZI. (c),(d) Correlation measurements performed with above-band CW excitation at $P_\mathrm{sat}/4$ and a $\delta\tau_p=22.9$\,ns delay. (e) Two-photon interference visibility calculated from (d). Black symbols: measured data. Blue curves: model fits described in text. Red curves: after convolution with detector response.}
    \label{Figure3}
\end{figure}

To measure TPI visibilities, the polarization of the $X^{1-}$ photons are first aligned to the slow axis of a polarization-maintaining (PM) fibre using a fibre paddle polarization controller and filtered using the tunable filter to isolate the $X^{1-}$ line. The photons are then input to a PM fibre-based unbalanced Mach-Zehnder interferometer (MZI) equipped with two fibre-coupled free-space non-polarizing beamsplitters (BS1 and BS2) each with a 50:50 nominal splitting ratio, see Fig.~\ref{Figure3}(a). Two additional polarizers are placed on each arm of the MZI to ensure linear polarized photons incident on BS2. One arm of the MZI includes a half-wave plate (HWP) for controlling the relative polarizations, $\phi$, of the photons incident on the input ports of BS2. The delay between the two arms of the interferometer, $\delta\tau_p$, is adjusted by adding additional fibre to one of the arms. Two SNSPD detectors at the output ports of BS2, labelled `start' and `stop', together with counting electronics are used to register coincidences. 

We discuss first measurements performed using CW excitation. We verified the performance of the interferometer by measuring the dependence of the TPI visibility at $\tau=0$\,ns on $\phi$, given by $V(0,\phi)=[g^{(2)}(0,\phi)-g^{(2)}(0,0^\circ)]/g^{(2)}(0,\phi)$. This is plotted in Fig.~\ref{Figure3}(b) (black  symbols), measured using above-band CW excitation and a delay of $\delta\tau_p=1.84$\,ns. We observe an expected oscillatory behaviour in the visibility, from $V \sim 0.9$ to $V \sim 0$, as $\phi$ is varied from $0^\circ$ to $\pm 45^\circ$. The maximum visibility observed is limited by the detector response, discussed below.

To determine the TPI visibilities as a function of delay $\tau$, coincidence spectra are measured for cross- ($\phi= 45^\circ$) and co-linear ($\phi=0^\circ$) polarized photons incident on BS2, $g_\perp^{(2)}(\tau)$ and $g_\parallel^{(2)}(\tau)$, respectively, and the visibilities are calculated from 

\begin{equation}
        V(\tau)=(g_\perp^{(2)}(\tau)-g_\parallel^{(2)}(\tau))/g_\perp^{(2)}(\tau). 
   \label{V}
\end{equation} Typical spectra are shown in Figs.~\ref{Figure3}(d) (black symbols) for measurements made using above-band CW excitation at $P_\mathrm{sat}/4$ and a delay $\delta\tau_p=22.9$\,ns. 

To model the coincidence spectra, we examine the four possible path combinations two photons can take to traverse the MZI and arrive simultaneously at BS2 at $\tau=0$. Consider first the 2 cases where the photons are incident on different input ports of BS2 (e.g. the photons travel in different arms of the MZI). If the pair is distinguishable (e.g. in the polarization degree of freedom by setting $\phi= 45^\circ$) each photon is equally likely to be reflected or transmitted, and of the four possible outcomes, only the two where the photons exit different ports will register coincidence counts at zero delay. This is the case in the upper panel of Fig.~\ref{Figure3}(d) which shows a $g_\perp^{(2)}(\tau=0) \sim 0.5$ relative to the coincidence counts at long delays from uncorrelated photons. If the pair is indistinguishable, these two outcomes will destructively interfere (i.e. the incident pair coalesces, always exiting the same port) so that there are no possible outcomes that can register a coincidence at zero delay. This is the case in the lower panel of Fig.~\ref{Figure3}(d) which shows a $g_\parallel^{(2)}(\tau=0) \sim 0$. The other two cases, where the photons travel in the same arm of the MZI, will have both photons incident on the same port of BS2. Since the photons are generated by the same single photon source, they cannot arrive simultaneously, hence will not register zero-delay coincidences. 

The absence of simultaneous photons incident on BS1 also eliminates one of the four possible outcomes that would lead to coincidences at $\tau=\pm\delta\tau_p$. This manifests as a reduction in $g^{(2)}(\tau)$ to $\sim0.75$ at $\tau=\pm\delta\tau_p$ that is seen in Fig.~\ref{Figure3}(d).


For 50:50 beamsplitters, the behaviour above can be modelled by

\begin{equation}
         g_\perp^{(2)}(\tau) = \frac{1}{2} g^{(2)}(\tau)+\frac{1}{4} [g^{(2)}(\tau-\delta\tau_{p})+g^{(2)}(\tau+\delta\tau_{p})]
    \label{eq:cw_HOM_perp_perf}
\end{equation}

\begin{equation}
    \begin{split}
           g_\parallel^{(2)}(\tau)= &\frac{1}{2} g^{(2)}(\tau)+\frac{1}{4}[g^{(2)}(\tau-\delta\tau_{p})+g^{(2)}(\tau+\delta\tau_{p})]\\
           &\times[1-Fe^{-2|\tau|/\tau_c^{'}}]
     \end{split}
    \label{eq:cw_HOM_parallel_perf}
\end{equation} which describe the cross-polarized, $g_\perp^{(2)}(\tau)$, and co-polarized, $g_\parallel^{(2)}(\tau)$, coincidences, respectively, in terms of the second-order correlation function $g^{(2)}(\tau)$. The first (second) term in Eqns.~\ref{eq:cw_HOM_perp_perf} and \ref{eq:cw_HOM_parallel_perf} accounts for the cases where the photons are incident on the same (different) input port(s) of BS2 whilst $F$ accounts for the spatial overlap of the photons on BS2 which is assumed to be 100\% (i.e. $F=1$). The time-scale $\tau^{'}_c$ represents, phenomenologically, the temporal extent over which photons incident on BS2 will coalesce, providing a measure of the probability of coalescence. The functional form with which it is incorporated in Eq.~\ref{eq:cw_HOM_parallel_perf} will depend on the mechanism limiting coalescence e.g. homogeneous versus inhomogeneous broadening. For simplicity, we use an exponential decay (i.e. the visibility is limited by pure dephasing\cite{Bylander_EPJD2003}) though we do expect spectral diffusion to play a role due to the non-resonant excitation.

In this work we use a modified model which explicitly includes the transmission ($T_\mathrm{BS1}$, $T_\mathrm{BS2}$) and reflection ($R_\mathrm{BS1}$, $R_\mathrm{BS2}$) coefficients of BS1 and BS2:\cite{Patel_PRL2008} 

\begin{equation}
    \begin{split}
        g_\perp^{(2)}(\tau) = &4(T_\mathrm{\tiny BS1}^2+R_\mathrm{BS1}^2)R_\mathrm{BS2}T_\mathrm{BS2}g^{(2)}(\tau)\\
        &+4R_\mathrm{BS1}T_\mathrm{BS1}[T_\mathrm{BS2}^2g^{(2)}(\tau-\delta\tau_{p})\\&+R_\mathrm{BS2}^2g^{(2)}(\tau+\delta\tau_{p})]
    \end{split}
    \label{eq:cw_HOM_perp}
\end{equation}

\begin{equation}
    \begin{split}
        g_\parallel^{(2)}(\tau)= &4(T_\mathrm{BS1}^2+R_\mathrm{BS1}^2)R_\mathrm{BS2}T_\mathrm{BS2}g^{(2)}(\tau)
        \\&+4R_\mathrm{BS1}T_\mathrm{BS1}[T_\mathrm{BS2}^2g^{(2)}(\tau-\delta\tau_{p})\\&+R_\mathrm{BS2}^2g^{(2)}(\tau+\delta\tau_{p})]
        \times [1-F e^{-2|\tau|/\tau^{'}_c}].
    \end{split}
    \label{eq:cw_HOM_parallel}
\end{equation}

The additional terms are required to account for deviations from a perfect, loss-less system, deviations which result in coincidence dip depths that differ from the values described above. For example, if T$_\mathrm{BS1}\neq R_\mathrm{BS1}$, then the drop in coincidences at $\tau=\pm\delta\tau_p$ will be less than 25\%, and, in the case of distinguishable photons, $g_\perp^{(2)}(\tau=0) < 0.5$. For $T_\mathrm{BS2} \neq R_\mathrm{BS2}$, there will be an asymmetry in the depths of the $\tau=\pm\delta\tau_p$ side dips. Although there is little evidence of the latter in the measured spectra (i.e. $T_\mathrm{BS2} \sim R_\mathrm{BS2}$), we do observe values of $g_\perp^{(2)}(0) < 0.5$ which we associate not necessarily to an unbalanced BS1, but to a difference in the propagation losses between the two arms of the MZI (e.g. from inclusion of the components and accompanying mating sleeves) that result in different count rates incident on the two ports of BS2.

To reproduce the experimental $g_\perp^{(2)}(\tau)$ and $g_\parallel^{(2)}(\tau)$ using Eqns.~\ref{eq:cw_HOM_perp} and \ref{eq:cw_HOM_parallel}, we first measure $g^{(2)}(\tau)$ by detecting coincidences at the two output ports of BS1. The $g^{(2)}(\tau)$ measured under the same operating conditions as the HOM experiment shown in Fig.~\ref{Figure3}(d) is shown in Fig.~\ref{Figure3}(c) (black symbols). The response is modelled using $g^{(2)}(\tau)=1-e^{-(1/T_1+R)|\tau|}$ (see Supplemental Material\cite{SM_2023}) where the radiative lifetime, $T_1=1.75$\,ns, is independently measured and the excitation rate, $R$, is a fit parameter. The blue curve in the figure shows the model fit using $R=0.1\,\mathrm{ns}^{-1}$.

This expression of $g^{(2)}(\tau)$ is incorporated in Eqns.~\ref{eq:cw_HOM_perp} and \ref{eq:cw_HOM_parallel} which are then applied to the measured correlations $g_\perp^{(2)}(\tau)$ and $g_\parallel^{(2)}(\tau)$ with $\tau_\mathrm{c}^{'}$ and the beamsplitter coefficients as fit parameters. The resulting fits are shown in Fig.~\ref{Figure3}(d) (blue curves) where, for this particular measurement (above-band CW excitation at $P_\mathrm{sat}/4$ and $\delta\tau_p=22.9$\,ns delay) we obtain a $T_\mathrm{BS1}:R_\mathrm{BS1}$ ratio of 0.25:0.75, a $T_\mathrm{BS2}:R_\mathrm{BS2}$ ratio of 0.48:0.52 and $\tau^{'}_c= 0.55$\,ns. 

The visibility obtained from Eqn.~\ref{V} using the model correlations is shown in Fig.~\ref{Figure3}(e) (blue curve) and, by definition, predicts a TPI visibility $V(\tau=0)=1$. To match experiment, we convolve the model correlations with a 100\,ps Gaussian detector response function  and obtain the red curves in Figs.~\ref{Figure3} (c)-(e) which correctly predict the measured raw visibility $V(\tau=0)\sim 0.85$.

\begin{figure}
    \includegraphics[width=8.cm,clip=true]{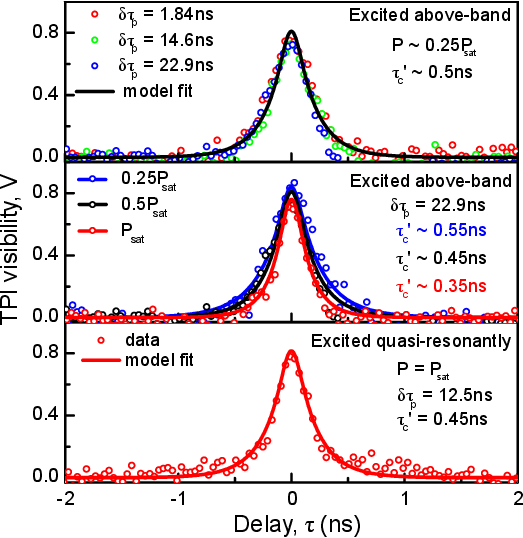}
    \caption{TPI visibilities obtained under different operating conditions. Upper panel: visibilities using different delays $\delta\tau_\mathrm{p}$. Middle panel: visibilities as a function of excitation power. Lower panel: visibility under quasi-resonant excitation into a p-shell of the dot.}
    \label{Figure4}
\end{figure}

We have performed the above analysis on measurements taken under different operating conditions to extract $\tau^{'}_c$ values. The results are summarized in Fig.~\ref{Figure4}. For measurements as a function of $\delta\tau_\mathrm{p}$ using above-band CW excitation at $P=0.25P_\mathrm{sat}$ (upper panel) we obtained $\tau^{'}_c \sim 0.5$\,ns, independent of path delay for $\delta\tau_\mathrm{p} = 1.84$\,ns to 22.9\,nm.  This suggests that the mechanisms limiting $\tau^{'}_c$ values occur on time-scales faster than $\sim 2$\,ns. In contrast, from measurements as a function of excitation power (middle panel), we observed a significant increase in $\tau^{'}_c$ from 0.35\,ns to 0.55\,ns for a four-fold reduction in power. We also observed a moderate increase in $\tau^{'}_c$ when we excite quasi-resonantly into a p-shell of the dot. For measurements at $P=P_\mathrm{sat}$, $\tau^{'}_c=0.45$\,ns for p-shell excitation (bottom panel) compared to $\tau^{'}_c=0.35$\,ns for above-band excitation.

Although CW HOM measurements reveal the presence of decoherence mechanisms through the measurement of $\tau^{'}_c$, to extract meaningful values requires simultaneous fitting of the three correlations $g^{(2)}(\tau)$, $g_\perp^{(2)}(\tau)$ and $g_\parallel^{(2)}(\tau)$ measured under the same operating conditions. Without the additional information provided by $g^{(2)}(\tau)$, erroneous values of $\tau^{'}_c$ are possible due the dependence of the antibunching dip in CW measurements on the excitation rate $R$, see, for example, Ref.~\citenum{Laferriere_APL2021} and Supplemental Material\cite{SM_2023}, Fig.~S1. Finally we note that in all cases $V(\tau=0)=1$ if account is taken of the detector response. This is expected in CW HOM measurements if $g^{(2)}(\tau=0)=0$, as is the case here i.e. $V(\tau=0)$ is only limited by the detector response\cite{Patel_PRL2008,Ates_PRL2009,Patel_NP2010}.

\section{TPI measurements: pulsed excitation}

\begin{figure}
    \includegraphics[width=8.cm,clip=true]{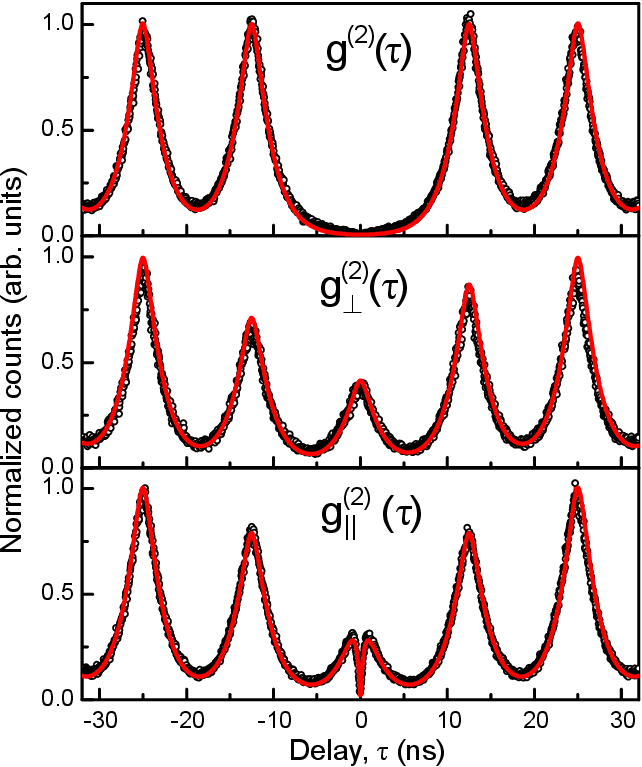}
    \caption{Correlation measurements for quasi-resonant pulsed excitation at $P_\mathrm{sat}/4$ (black symbols). Red curves are model fits described in the text.}
    \label{Figure5}
\end{figure}

To evaluate non-post-selected visibilities e.g. over the temporal extent of the emitted photons, we perform the same TPI measurements using pulsed excitation which allows one to quantify the probability of emitting identical photons and will also be ultimately required for on-demand operation. Using a tunable pulsed laser (pulse width $\sim 2$\,ps) we measure visibilities as in the previous section, for both above-band and quasi-resonant excitation. We use a pulse repetition rate of 80\,MHz, i.e. a pulse period of 12.5\,ns, and a corresponding delay in one arm of the MZI of $\delta\tau_{p}=12.5$\,ns. The measured coincidences $g^{(2)}(\tau)$, $g_\perp^{(2)}(\tau)$ and $g_\parallel^{(2)}(\tau)$ (black symbols) for the case of quasi-resonant excitation at $P_\mathrm{sat}/4$ are shown in Fig.~\ref{Figure5}.

Similar to the case of CW excitation for a nominal interferometer (50:50 beamsplitters, 100\% transmission), if two photons arriving simultaneously at BS2 are distinguishable we expect a peak centred at zero delay having half the height of the peaks at $\pm 2T$ whilst the peaks at $\pm T$ should be reduced by 25\%. For two perfectly indistinguishable photons arriving at BS2, the zero-delay peak should be absent. The measured $g_\perp^{(2)}(\tau)$ (middle panel in figure) qualitatively reproduces the behaviour expected of impinging distinguishable photons on BS2. For the indistinguishable case, however, the zero-delay peak in the measured  $g_\parallel^{(2)}(\tau)$ (bottom panel in figure) is still present but with a dip that drops to zero coincidences at $\tau=0$.  This behaviour is well documented\cite{Flagg_PRL2010,Gold_PRB2014,Huber_NJP2015,Reimer_PRB2016,Weber_PRB2018} and is a consequence of processes that limit coalescence to time-scales shorter than the temporal extent of the photons e.g.  ($i$) pure dephasing\cite{Legero_APB2003}, ($ii$)  spectral diffusion\cite{Kambs_NJP2018} and ($iii$) excitation timing jitter\cite{Kiraz_PRA2004}, and thus limit extracted $\tau^{'}_c$ to values less then $2T_1$. We note that the last mechanism, timing jitter, is the primary distinction between the pulsed and CW measurements: it is absent in the latter where the experiment selects only the photons that arrive simultaneously at BS2. 

As in the CW case, the curves in Fig.~\ref{Figure5} are modelled using Eqns.~\ref{eq:cw_HOM_perp} and \ref{eq:cw_HOM_parallel} but with the latter modified to\cite{SM_2023}: 

\begin{equation}
    \begin{split}
        g_\parallel^{(2)}(\tau)= &4(T_\mathrm{BS1}^2+R_\mathrm{BS1}^2)R_\mathrm{BS2}T_\mathrm{BS2}g^{(2)}(\tau)
        \\&+4R_\mathrm{BS1}T_\mathrm{BS1}[(T_\mathrm{BS2}^2g^{(2)}(\tau-\delta\tau_{p})
        \\&+R_\mathrm{BS2}^2g^{(2)}(\tau+\delta\tau_{p}))
        \\&-(T_\mathrm{BS2}^2g^{(2)}(-\delta\tau_{p})+R_\mathrm{BS2}^2g^{(2)}(+\delta\tau_{p}))F e^{-2|\tau|/\tau^{'}_c}]
            \end{split}
    \label{eq:pulsed_HOM_parallel}
\end{equation} such that $g_\parallel^{(2)}(\tau)$ is simply given by $g_\perp^{(2)}(\tau)$ less the exponential term defining the time-scale over which the photons coalesce. This is seen clearly for the case of nominal 50:50 beamsplitters in which case Eq.~\ref{eq:pulsed_HOM_parallel} reduces to $g_\perp^{(2)}(\tau)-0.5Fe^{-2|\tau|/\tau^{'}_c}$ after setting $g^{(2)}(-\delta\tau_{p})=g^{(2)}(+\delta\tau_{p})=1$.

The $g^{(2)}(\tau)$ that is incorporated in Eqs.~\ref{eq:cw_HOM_perp} and \ref{eq:pulsed_HOM_parallel} is constructed from peaks described by $e^{(-\tau/T_1)}[1-e^{(-\tau/\tau_e)}] \ast e^{(\tau/T_1)}[1-e^{(\tau/\tau_e)}]$, i.e. we self-convolve model fits to measured time-resolved PL spectra, see Supplemental Material\cite{SM_2023}, Fig.~S2. This allows us to account for the dependence of the $g^{(2)}(\tau)$ peaks on operating conditions through $\tau_e$, the time-scale associated with preparation of the $X^{1-}$ state. We neglect re-excitation which reduces the single photon purity to $\sim 98\%$ and assume a $g^{(2)}(\tau)=0$ in the zero delay peak. The resulting fits are shown in the figure (red curves) where, as for the CW measurements, we have fit the beam splitter ratios and $\tau^{'}_c$. We note that for the pulsed measurements we obtained slightly different beam splitter ratios for cross- and co-linear measurements which is evident in the data from the pronounced asymmetry between the peaks at $\tau=\pm T$ in the $g_\perp^{(2)}(\tau)$ whereas these peaks are symmetric in $g_\parallel^{(2)}(\tau)$ correlations. For the particular measurement shown in Fig.~\ref{Figure5}, i.e. quasi-resonant excitation at $P_\mathrm{sat}/4$, we obtain a $T_\mathrm{BS1}:R_\mathrm{BS1}$ ratio of 0.27:0.73 (0.31:0.69) and  $T_\mathrm{BS2}:R_\mathrm{BS2}$ ratio of 0.35:0.65 (0.5:0.5) for $g_\perp^{(2)}(\tau)$ ($g_\parallel^{(2)}(\tau)$) and $\tau^{'}_c = 0.95$\,ns. 

\begin{figure}
    \includegraphics[width=8.5cm,clip=true]{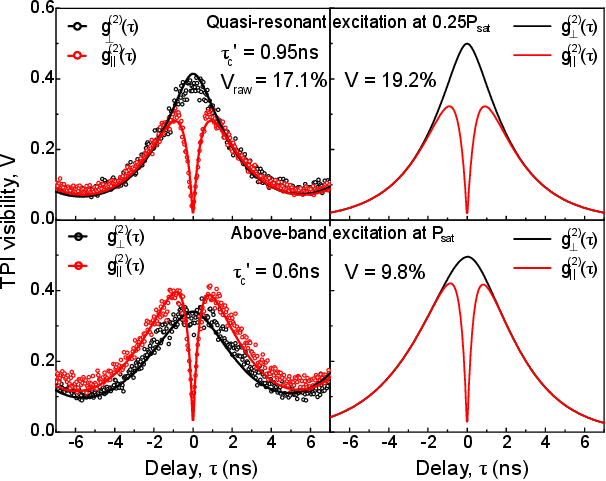}
    \caption{Zero-delay correlation peaks under quasi-resonant (upper panel) and above-band (lower panel) excitation. Left panel shows the raw data (symbols) and model fits (curves). Right panel shows model calculations with BS1 and BS2 ratios set to 50:50 and contributions from side peaks removed.}
    \label{Figure6}
\end{figure}

In Fig.~\ref{Figure6} we plot a zoom-in of the the zero-delay peaks for both cross- and co-linear measurements, $g_\perp^{(2)}(\tau)$ and $g_\parallel^{(2)}(\tau)$ respectively, for quasi-resonant (upper panel) and above-band (lower panel) excitation. The raw visibility over the temporal extent of the emitted photons is determined as in the previous section using Eqn.~\ref{V} but here we use the integrated coincidence counts over $\tau \pm T/2$. Under quasi-resonant excitation at $P=P_\mathrm{sat}/4$ we obtain a non-post-selected visibility of $V=17.1\%$. If the contributions to the coincidence counts from the correlation side peaks are removed (see right panel of figure), we obtain a corrected visibility (e.g. the visibility that would result using a longer pulse period $T$) of $V=19.2\%$. In this plot we have also corrected for non-nominal values of the beam splitter ratios for comparison with the above-band measurements, discussed below. 

For above-band excitation (lower panel of Fig.~\ref{Figure6}) the zero-delay peak is significantly broader than that observed when exciting quasi-resonantly (e.g. compare $g_\perp^{(2)}(\tau)$ (black curves)) and the dip at $\tau=0$ is significantly narrower (compare $g_\parallel^{(2)}(\tau)$ (red curves)). Both broadening of the peak and narrowing of the dip (quantified by the reduced $\tau^{'}_c=0.6$\,ns) are a consequence of the more significant timing jitter\cite{Legero_AAMOP2003} present in above-band excitation i.e. longer time-scale associated with the state preparation $\tau_e$, since above-band excitation includes additional processes related to carrier thermalization and capture not present for quasi-resonant excitation. We note that here we have used a higher excitation power than that used in the quasi-resonant measurement and this will also contribute to a decrease in $\tau^{'}_c$, as observed in the CW measurements. 

We also observe that for the above-band measurements the zero-delay peaks of cross- and co-linear correlations do not overlap at $\tau$ values away from $\tau=0$. This is a consequence of the more significant difference in the beam splitter ratios between the respective measurements and strongly impacts the calculated visibility using the raw data. Instead we calculate $g_\perp^{(2)}(\tau)$ and $g_\parallel^{(2)}(\tau)$ using the fit value of $\tau^{'}_c$ but with 50:50 beam splitter ratios as above. For the above-band non-post-selected visibility we obtain $V=9.2\%$, substantially lower compared to the quasi-resonant case and consistent with the reduced $\tau^{'}$ value. 

Comparison with previously reported values is restricted due to the dependence of measured visibilities on both excitation conditions \cite{Huber_NJP2015,Reindl_PRB2019} and the delay between interfered photons,  $\delta\tau_{p}$\cite{Thoma_PRL2016}. For above-band excitation, the only reported values, to our knowledge, are from our previous work\cite{Reimer_PRB2016} where visibilities of $V\sim 5\%$ were measured. There, a temporal delay of $\delta\tau_{p} = 50$\,ns was used and the devices were grown at lower temperature where more severe linewidth broadening is expected, see Ref.~\citenum{Laferriere_PRB2023}. In the case of p-shell excitation with $\delta\tau_{p} \sim 12.5$\,ns, similar visibilities, in the range $V = 0.21 - 0.59$, have been previously reported \cite{Gschrey_NC2015,Thoma_PRL2016,Wang_PRL2016} including from on-chip sources\cite{Kirsanske_PRB2017}.


\section{Coherence measurements}\label{coherence}

In this section we determine the coherence properties of the two-level excitonic system. We show results in the time-domain where coherence times, $\tau_c$, are extracted from single photon interference visibilities e.g. $g^{(1)}(\tau)$ measurements. We also compare with results in the frequency domain, where coherence times are extracted from linewidth measurements.

\subsection{Interferometric measurements}

\begin{figure}
    \includegraphics[width=8.cm,clip=true]{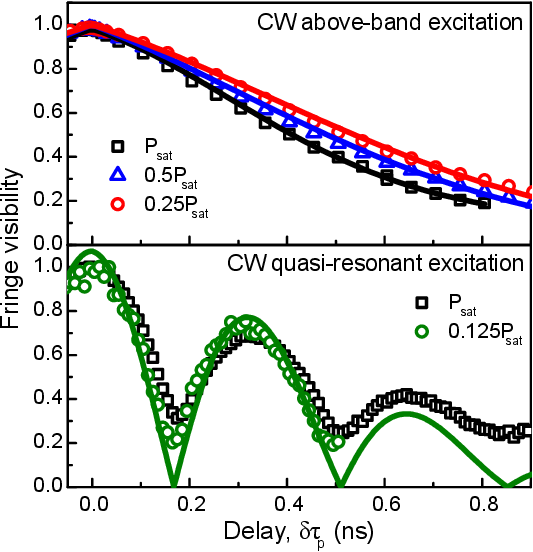}
    \caption{Fringe visibility extracted from single-photon interference measurements using above-band (upper panel) and quasi-resonant (lower panel) excitation. Curves are model fits described in the text.}
    \label{Figure7}
\end{figure}

For the time-domain measurements, the MZI was balanced (e.g. $\delta\tau_{p}=0$) and a motorized fibre-based delay stage with a tuning range of 1.2\,ns was added to one arm. The stage was scanned across it's full range and at selected delays, $\delta\tau_{p}$, the fringe visibility was determined using a phase modulator in one arm of the MZI. The fringe visibility as a function of $\delta\tau_{p}$ extracted from above-band and quasi-resonant measurements are plotted in Fig.~\ref{Figure7}.

For a homogeneously broadened transition i.e. the spectral linewidth of the emitted photons corresponds to the natural linewidth, the lineshape is Lorentzian and the visibility is expected to decay exponentially with a time constant $T_2$. In the presence of inhomogeneous broadening, the decay will have a Gaussian component\cite{Berthelot_NP2006}, $T_G$, and is more appropriately described by a Voigt profile\cite{Reimer_PRB2016}:

\begin{equation}
    g^{(1)}(\delta\tau_p)\sim \exp \left[-\frac{\pi}{2}\left(\frac{\delta\tau_p}{T_\mathrm{G}}\right)^2-\frac{|\delta\tau_p|}{T_2}\right].
\label{g1_voigt}
\end{equation}

We model the above-band fringe visibility using Eqn.~\ref{g1_voigt} (curves in the upper panel of Fig.~\ref{Figure7}) to extract $T_2$ and $T_\mathrm{G}$. To compare with linewidth measurements below, we also calculated the full width at half maximum (FWHM) of the Voigt profile in the frequency domain given by 

\begin{equation}
    \delta \omega_\mathrm{V} = 0.535\delta \omega_\mathrm{L} + \sqrt{0.217\delta \omega^2_\mathrm{L}+\delta \omega^2_\mathrm{G}}.
\label{omega_V}
\end{equation} where $\delta \omega_\mathrm{L} = \frac{1}{\pi T_2}$ is the Lorentzian contribution to the linewidth and $\delta \omega_\mathrm{G} = \frac{\sqrt{2\ln2}}{\sqrt{\pi}T_\mathrm{G}}$ is the Gaussian contribution. The extracted Voigt linewidths, $\delta\omega_V$, using above-band excitation are plotted in Fig.~\ref{Figure9} (filled symbols).

In the case of the quasi-resonant measurements (lower panel of Fig.~\ref{Figure7}), instead of a simple decay in the visibility with $\delta\tau_p$, we observe oscillatory behaviour indicative of a beating phenomenon. Why this arises is discussed below; here we simply assume the presence of two lines with frequency separation $\omega_s$ and identical coherence and model this behaviour by multiplying Eqn.~\ref{g1_voigt} by the beating term $|\cos(\omega_s\delta\tau_p)|$. We only model the measurement at $P=0.125P_\mathrm{sat}$: at higher excitation powers (e.g. $P=P_\mathrm{sat}$) the assumption of identical coherence properties likely does not apply as suggested by the higher minima values in the visibility that are observed in the figure. The extracted $\delta\omega_V$ from Eqn.~\ref{omega_V} from the quasi-resonant pump at $P=0.125P_\mathrm{sat}$ is plotted in Fig.~\ref{Figure9} (open symbol).

\subsection{Linewidth measurements}

\begin{figure}
    \includegraphics[width=8.cm,clip=true]{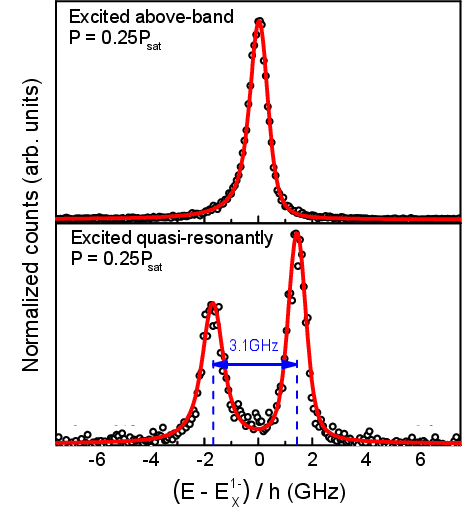}
    \caption{High-resolution photoluminescence spectrum of the $X^{1-}$ emission for CW above-band excitation (upper panel) and CW quasi-resonant excitation (lower panel) at $P_\mathrm{sat}/4$. Red curves are Voigt fits to the data.}
    \label{Figure8}
\end{figure}

In the frequency-domain, we measure the linewidth of the emitted photons using a fibre-based, piezo-controlled Fabry-Perot etalon (bandwidth: 250\,MHz, free spectral range: 40.75\,GHz). High-resolution spectra obtained by scanning the etalon through the $X^{1-}$ emission peak are shown in Fig.~\ref{Figure8} for CW above-band (upper panel) and quasi-resonant (lower panel) excitation at $P_\mathrm{sat}/4$. For above-band excitation we observe a single peak as expected from a singly-charged complex whilst for quasi-resonant excitation the same emission line is a doublet with a splitting of $\omega_s/\pi= 3.1$\,GHz consistent with the beating frequency observed in the coherence measurements. Given that $g^2(0) \sim 0$, the observation of two peaks suggests either two mutually exclusive excitonic complexes or the same excitonic complex with the electrostatic environment jumping between two possible states.

There are two possible reasons why the observation of either two complexes or the same complex in two environments would depend on the excitation energy. First, in quasi-resonant excitation, fewer carriers are introduced into the system, resulting in a different Fermi level profile compared to above-band excitation. This may modify the relative intensities of different charge complexes in time-integrated PL spectra, which we have previously observed. Second, there may be a single, defect-related trap sufficiently close to the dot such that a Stark-mediated shift of the $X^{-1}$ emission energy of $\omega_s\sim 3$\,GHz will result depending on the occupation of the trap. In this scenario, for above-band excitation the trap occupation is constant on the time-scale of the measurement ($\sim 100$\,ms) whereas for quasi-resonant excitation, the occupation fluctuates on a much faster time-scale such that both emission energies are observed in the time-integrated PL. 

The magnitude of the observed splitting depends on the quasi-resonant excitation power, decreasing by $\sim 0.4$\,GHz as the power is increased from $0.25P_\mathrm{sat}$ to $P_\mathrm{sat}$. Such an excitation power-dependent splitting is consistent with carrier screening of the Stark field but difficult to explain in the case of two distinct complexes. We therefore attribute the two peaks observed when exciting quasi-resonantly to the same excitonic complex $X^{1-}$.

\begin{figure}
    \includegraphics[width=8.5cm,clip=true]{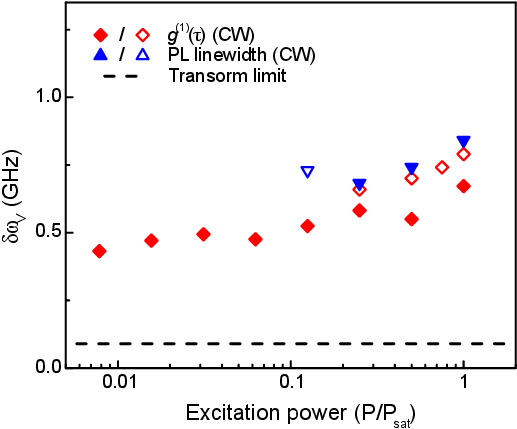}
    \caption{Voigt linewidths extracted from $g^{(1)}(\tau)$ and high resolution PL measurements as a function of excitation power. Filled (open) symbols correspond to above-band (quasi-resonant) excitation. Dashed horizontal line indicates the transform-limited linewidth, $\delta \omega = 1/2\pi T_1$.}
    \label{Figure9}
\end{figure}

High resolution spectra were measured as a function of excitation power, deconvolved from the etalon response and fit using a Voigt lineshape. We note that the lineshapes obtained using above-band excitation present a slight asymmetry when excited at higher powers and these were fit with an asymmetric Voigt function. The origin of this asymmetry is unclear, but is typical of the nanowire quantum dot system\cite{Laferriere_PRB2023}. The total linewidth $\delta\omega_V$ from the fits are plotted in Fig.~\ref{Figure9} for both the above-band (filled) and quasi-resonant (open) excitation. For clarity, only measurements of one of the two peaks observed with quasi-resonant excitation are included (the power-dependent linewidths of the second peak are similar).

Comparison of the time- and frequency domain measurements reveal similar linewidths which do not depend significantly on the excitation energy: above-band (filled symbols) versus quasi-resonant (open symbols). The absence of any narrowing when exciting below-band is surprising: in fact, we observe a small increase in the measured linewidth for quasi-resonant excitation. This suggests that the mechanisms responsible for the excess broadening are unrelated to the much higher density of excess carriers or phonons introduced with above-band excitation. 

In all cases, we observe a decrease in excess broadening at lower excitation powers with a minimum extracted linewidth of 4X the transform limit (dashed line in the figure). The reduction in linewidth with excitation power is expected from a reduction in both phonon-related pure dephasing and charge noise-related spectral diffusion. We do not attempt to quantify these two contributions in terms of the Lorentzian versus Gaussian contributions to the measured linewidth: although the total Voigt linewidth is considered accurate, the Lorentzian:Gaussian ratio likely has a large uncertainty.

\section{Discussion}

\begin{figure}
   \includegraphics[width=8.5cm,clip=true]{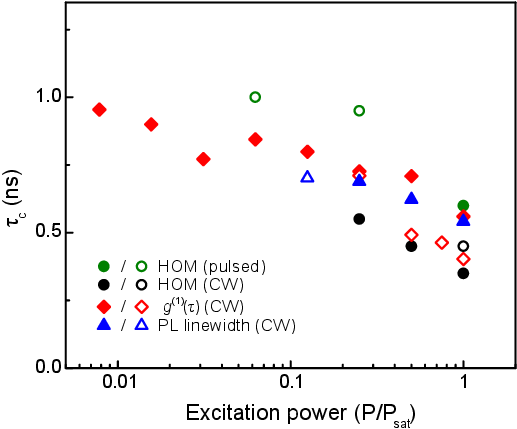}
   \caption{Coherence times extracted from HOM, $g^{(1)}(\tau)$ and high resolution PL measurements as a function of excitation power. Filled (open) symbols correspond to above-band (quasi-resonant) excitation.}
    \label{Figure10}
\end{figure}

To compare the coherence measurements with the HOM results, we use $T_L$ and $T_G$ extracted from the $g^{(1)}(\tau)$ and linewidth measurements to calculate a `coherence time' $\tau_c$ given by\cite{Kambs_NJP2018}

\begin{equation}
    \tau_c=-\frac{T_G^2}{\pi T_2} + \sqrt{\left(\frac{T_G^2}{\pi T_2}\right)^2+\frac{2T_G^2}{\pi}}.
    \label{tau_c}
\end{equation} The calculated values of $\tau_c$ from the three experiments are summarized in Fig.~\ref{Figure10}. All measurements give $\tau_c$ values of $0.5 \pm 0.1$\,ns at $P=P_\mathrm{sat}$ which increase as the excitation power is reduced, consistent with the linewidth measurements in Fig.~\ref{Figure9}, but still well below the transform limit of $2T_1$.

In the following we consider the distinct nature of each experiment to extract information on the nature of the mechanisms responsible for $\tau_c<2T_1$. We consider first the different time scales associated with each experiment. In the HOM measurement, different photons are interfered on a time-scale of 2-23\,ns whilst in the coherence measurement, interference is between the same photon and the time-scale is 1.2\,ns, but each point in the fringe visibility trace is a statistical average over $\sim 100$\,ms. In the linewidth measurements, on the other hand, there is no interference, only a measure of the spectral purity over time-scales in the seconds. The consistency in the extracted coherence times irrespective of the experiment thus suggests that the mechanism(s) responsible for reducing values of $\tau_c$ below the transform limit of $2T_1$ occur on a timescale of $< 2$\,ns and there are no other mechanisms on longer time-scales for at least seconds. 

Next we note the absence of any significant improvement in $\tau_c$ when switching from above-band to quasi-resonant excitation which is in stark contrast to other studies\cite{Huber_NJP2015}. This suggests that phonon dephasing is not the primary mechanism limiting the coherence times since the phonon occupation, by necessity, will be higher for above-band excitation\cite{Laferriere_PRB2023}. Which leaves spectral diffusion as the likely source, meaning the charge environment is equally stable, regardless of excitation mode.

Finally, we note that all the measurements with the exception of the pulsed HOM experiments, are independent of excitation jitter. And yet it is these measurements that produced the highest values of $\tau_c$ (compare, for example, pulsed versus CW HOM in Fig.~\ref{Figure10}). This suggest that either timing jitter is relatively unimportant or that pulsed excitation produces a more stable charge environment thus compensating for any reduction in $\tau_c$ due to timing jitter.

\section{Conclusion}

In summary, we have demonstrated the generation of indistinguishable photons on-chip based on the hybrid integration of positioned-controlled single quantum dot nanowires and silicon-based photonic integrated circuitry\cite{Mnaymneh_AQT2020}. Non-post-selected measurements over the photon lifetime revealed coalescence of only a small fraction of the emitted photons, $\sim 20\%$. The TPI visibility was limited by the excess broadening that arises when exciting non-resonantly. Higher visibilities are anticipated with coherent excitation, as demonstrated in other quantum dot systems\cite{He_NN2013,Ding_PRL2016,Somaschi_NP2016}.

The described integration approach provides a route to developing a platform whereby multiple sources of indistinguishable photons can be selectively incorporated on-chip, a long-term goal of future quantum technologies. As a final note, in this work we investigated nanowire quantum dots emitting at $\lambda \sim 980$\,nm. However, the InAs/InP material system is also the ideal choice for generating telecom wavelengths and we have recently demonstrated high quality sources emitting in the O-band\cite{Laferriere_NL2023}. Using such emitters, the pick and place approach described here can be applied to the highly developed silicon-on-insulator integrated photonics platform.


\section*{Acknowledgments}

This work was supported by the Natural Sciences and Engineering Research Council of Canada through the Discovery Grant SNQLS, by the National Research Council of Canada through the Small Teams Ideation Program QPIC and by the Canadian Space Agency through a collaborative project entitled ‘Field Deployable Single Photon Emitters for Quantum Secured Communications'. The authors would also like to thank Khaled Mnaymneh for assistance with the commercial foundry tape-out.

\section*{DATA AVAILABILITY}
The data that support the findings of this study are available from the corresponding author upon reasonable request.

\section*{REFERENCES}

\bibliography{whiskers}   

\end{document}


\author{Edith Yeung}
\affiliation{National Research Council Canada, Ottawa, Ontario, Canada, K1A 0R6.}
\affiliation{University of Ottawa, Ottawa, Ontario, Canada, K1N 6N5.}
\author{David B. Northeast}
\affiliation{National Research Council Canada, Ottawa, Ontario, Canada, K1A 0R6.}
\author{Jeongwan Jin}
\affiliation{National Research Council Canada, Ottawa, Ontario, Canada, K1A 0R6.}
\author{Patrick Laferri{\`e}re}
\affiliation{National Research Council Canada, Ottawa, Ontario, Canada, K1A 0R6.}
\author{Marek Korkusinski}
\affiliation{National Research Council Canada, Ottawa, Ontario, Canada, K1A 0R6.}
\affiliation{University of Ottawa, Ottawa, Ontario, Canada, K1N 6N5.}
\author{Philip J. Poole}
\affiliation{National Research Council Canada, Ottawa, Ontario, Canada, K1A 0R6.}
\author{Robin L. Williams}
\affiliation{National Research Council Canada, Ottawa, Ontario, Canada, K1A 0R6.}
\author{Dan Dalacu}
\affiliation{National Research Council Canada, Ottawa, Ontario, Canada, K1A 0R6.}
\affiliation{University of Ottawa, Ottawa, Ontario, Canada, K1N 6N5.}

\title{On-chip indistinguishable photons using III-V nanowire/SiN hybrid integration: Supplementary Material}
\maketitle

\section{Theory}\label{theory}

In this section we derive the relevant formulas used to calculated the  coincidences measured in the CW and pulsed Hong-Ou-Mandel experiment. 
We begin by calculating systematically the second-order correlation
functions $g^{(2)}$ for the CW and pulsed excitation schemes.
Further, we develop a systematic treatment of the HOM visibility by
relating it to the dynamics of the quantum dot by the quantum regression 
theorem.
We then present the formulas $g_{||}^{(2)}(\tau)$ for the HOM visibility
in each excitation scheme.


\subsection{Second-order correlation under CW excitation}
\label{secCW}
In the continuous-wave excitation scheme, we assume that we deal with
a strictly two-level system excited non-resonantly (incoherently).
The higher-energy state, $\ket{1}$, corresponds to the exciton $X^{1-}$ trapped 
in the quantum dot, whilst the lower-energy state, $\ket{0}$, corresponds
to the final-state electron.
The state $\ket{1}$ is re-excited at the rate $R$ and recombines radiatively
at the rate $\frac{1}{T_1}$, expressed by the radiative lifetime $T_1$.
We formulate the rate equations for the time-dependent occupation probabilities 
$p_1(t)$ and $p_0(t)$ of the two levels in the following form:
\begin{eqnarray}
    \frac{dp_1(t)}{dt} &=& -\frac{1}{T_1} p_1(t) + R p_0(t),\label{rate1}\\
    \frac{dp_0(t)}{dt} &=& \frac{1}{T_1} p_1(t) - R p_0(t).
    \label{rate2}
\end{eqnarray}
The normalization condition $p_1(t)+p_0(t)=1$ at any time $t$.
By adding the two equations we find that this condition is naturally
satisfied.
We focus on the first equation and substitute $p_0(t)=1-p_1(t)$, obtaining
\begin{equation}
        \frac{dp_1(t)}{dt} = -\left(\frac{1}{T_1} + R\right)p_1(t) +R.
\end{equation}
Postulating the solution in the form
\begin{equation}
    p_1(t)=A(t)\exp\left[ -\left(\frac{1}{T_1} + R\right)t \right],
\end{equation}
after elementary calculus we obtain the prefactor $A(t)$
in the form:
\begin{equation}
    A(t) = \frac{T_1 R}{1 + T_1 R}\exp\left[ +\left(\frac{1}{T_1} + R\right)t \right] + A_0,
\end{equation}
where $A_0$ is derived from initial conditions.
In order to calculate the correlation function $g^{(2)}(t)$, we assume that
at $t=0$ a photon has been emitted, and as a result our system is necessarily
in the state $\ket{0}$ and begins to be repopulated.
The probability of detecting another photon at time $t$ is proportional to $p_1(t)$
with the above condition.
By normalizing it to the probability at $t\rightarrow\infty$ we obtain $g^{(2)}(t)$.
We therefore need to calculate the probability $p_1(t)$ with the condition
that $p_1(t=0)=0$, which determines our choice of $A_0$.
Elementary calculations give 
\begin{eqnarray}
    p_1(t) &=& \frac{T_1 R}{1+T_1 R}\left\{ 1 -
    \exp\left[ -\left(\frac{1}{T_1} + R\right)t \right]\right\},\\
    g^{(2)}(t) &=& \frac{p_1(t)}{p_1(t \rightarrow \infty)} =
    1 - \exp\left[ -\left(\frac{1}{T_1} + R\right)t \right].
    \label{g2_CW}
\end{eqnarray}
This formula, generalized to account for negative delays, is used in the main text.

\subsection{Second-order correlation under pulsed excitation}
\label{secPulsed}
We now develop the formula for the second-order correlation function 
$g^{(2)}(t)$ under the above-bandgap pulsed excitation scheme.
We continue to model the dynamics of the charged exciton in the quantum dot
by the two-level system of states $\ket{1}$ (charged exciton present) 
and $\ket{0}$ (final-state electron), with the radiative 
recombination rate $1/T_1$ related to the radiative lifetime $T_1$.
We assume that at time $t=0$ a strong pulse has been applied nonresonantly,
creating a large number $N(t)$ of electron-hole pairs in the sample.
The population $N(t)$ is time-dependent since the pairs decay, e.g., by
radiative or non-radiative recombination or diffusion.
If we characterize the decay process by the time constant $T_D$, then
\begin{equation}
    N(t) = N_0 \exp \left( -\frac{t}{T_D} \right),
\end{equation}
where $N_0$ is a constant factor dependent on the power of the excitation pulse
(we assume that all electron-hole pairs are created instantaneously).
The quantum dot is then refilled from this above-bandgap reservoir
at the average rate of $1/\tau_e$ per pair.
Therefore, the quantum dot reexcitation rate is
\begin{equation}
    R(t) = \frac{N_0}{\tau_e} \exp \left( -\frac{t}{T_D} \right).
\end{equation}
Here we assume that the capture of one electron-hole pair by the dot 
alters the population $N(t)$ to a negligible degree.
The above time-dependent refill rate can now be inserted into the rate equations
(\ref{rate1}) and (\ref{rate2}) to obtain the complete time-dependent
probability $p_1(t)$ proportional to the photon emission probability.
Through the factor $N_0$ this probability is pulse power dependent.
The solution can be obtained in a closed, semi-analytical form, however
we find it unwieldy for the analysis in the main text.
For any excitation power, we normalize the solution and
approximate it with a biexponential formula, which for the $n$-th 
pulse takes the form
\begin{equation}
    p_1^{(n)}(t) = P H(t-nT) \left[ 1 - \exp\left(-\frac{t-nT}{\tau_e}\right)
    \right]    
   \exp\left( -\frac{t-nT}{T_1}\right),
\end{equation}
with the pulses separated by the fixed time interval $T$, and $H(t)$ 
being the Heaviside step function.
The factor $P$ is calibrated so that there is only one photon per each 
pulse, according to the assumptions explained in the main text.
We see that, within each pulse, the short-time behavior of this 
emission probability is described by the reexcitation rate $1/\tau_e$,
while the long-time behavior is determined by the radiative lifetime 
$T_1$.
However, in this simple phenomenological approach we lose the
dependence on excitation power.
We can now express the probability of photon emission as a function of 
time for a train of pulses simply as a superposition
\begin{equation}
    p_1(t) =  \sum_n p_1^{(n)}(t).
\end{equation}

The {\em instantaneous} second-order unnormalized correlation 
function $g^{(2)}(t_1,t_2)$ expresses the probability of detecting a 
photon at time $t_2$ on condition that a photon has been detected at 
time $t_1$.
Let us denote the delay $\tau = t_2-t_1$ and assume that the photon
at time $t_1$ originated from the pulse with index $n=0$.
With these assumptions, and taking into account that only one photon
can be emitted as a result of each pulse, we have
\begin{equation}
    g^{(2)}(t_1,\tau) = p_1^{(0)}(t_1) \sum_{n\neq 0} 
    p_1^{(n)}(t_1 + \tau).
    \label{pulsed_inst_g2}
\end{equation}
The statistical analysis of the experimental data involves
performing an ensemble average of the above function.
We can calculate this ensemble average by integrating
\begin{equation}
    g^{(2)}(\tau) = \frac{1}{T} \int_{-\infty}^{\infty} g^{(2)}(t_1,\tau) dt_1.
\end{equation}
With our biexponential probability functions we arrive at an analytical 
formula
\begin{eqnarray}
    g^{(2)}(\tau) &=& \left( \frac{T_1}{2T} - 
    \frac{1}{T}\frac{\tau_e T_1}{2\tau_e + T_1} \right)    
    \sum_{n\neq 0} \exp\left(  
    -\frac{|\tau-nT|}{T_1} \right) \nonumber \\
    &+& \left( \frac{1}{2T}\frac{\tau_e T_1}{\tau_e + T_1}
    - \frac{1}{T}\frac{\tau_e T_1}{2\tau_e + T_1}\right)
    \sum_{n\neq 0} \exp\left(  
    - \frac{\tau_e + T_1}{\tau_e T_1}|\tau-nT| \right).
    \label{pulsed_average_g2}
\end{eqnarray}
The result (up to a constant factor) is identical to the one obtained by
taking the self-convolution of the probability function and 
removing the part arising 
from the central peak to account for the single photon per peak emission.
This latter procedure is mentioned in the main text.

\subsection{A general theoretical approach to two-photon
HOM interference visibility}

In this Section we present the general approach to calculating the two-
photon HOM interference visibility based on the quantum 
regression theorem.
We follow the method described by Kiraz et al.in 
Ref.~\onlinecite{Kiraz_PRA2004}
with modifications to the Bloch equations stemming from the specifics of
our system.

We begin with the schematic diagram of the experimental system
presented in Fig. 3(a) of the main text.
If we denote by indices $3$ and $4$ the output ports of the
beamsplitter BS2, then the (unnormalized) HOM cross-correlation 
function is given by:
\begin{equation}
    g_{||}^{(2)}(t,\tau) = \langle a_3^+(t) a_4^+(t+\tau)
    a_4(t+\tau) a_3(t)\rangle,
\end{equation}
where $a_m^+(t)$ ($a_m(t)$) is the creation (annihilation)
operator of a photon at time $t$ in the mode (channel) $m$, and
the triangular brackets denote the expectation value.
To make the HOM interference effects clear, one now expresses the
photon creation and annihilation operators at the output ports $3$, $4$
in terms of those at the input ports $1$, $2$ of the beamsplitter BS2.
For simplicity of presentation we assume BS2 to be a perfect $50$\%
-- $50$\% device; the approach can be readily generalized to any
transmission and reflection coefficients, as shown in the main text.
We have therefore
\begin{equation}
    g_{||}^{(2)}(t,\tau) = 
   \frac{1}{2} \langle a_1^+(t) a_2^+(t+\tau)a_2(t+\tau) a_1(t)\rangle -
   \frac{1}{2} \langle a_1^+(t) a_2^+(t+\tau)a_1(t+\tau) 
   a_2(t)\rangle,    
\end{equation}
with the crucial HOM interference term appearing with the negative sign.
Next, we express the photon operators via the quantum dot raising and
lowering (dipole) operators $\pi^+$ and $\pi$ [Ref.~\onlinecite{Loudon1983}].
The operators work on the quantum dot states in the following manner:
$\pi^+ \ket{0} = \ket{1}$ and $\pi \ket{1} = \ket{0}$,
with the state $\ket{1}$ ($\ket{0}$) correspond to the
confined charged exciton, or "filled dot" (the final state electron,
or "empty dot").
The equivalence between the photon and quantum dot operators is 
expressed as~\cite{Kiraz_PRA2004,Loudon1983}
\begin{equation}
    a(t) = A(\vec{r}) \pi\left(t - \frac{\vec{r}}{c}\right),
\end{equation}
where $A(\vec{r})$ is the time-independent proportionality factor,
$c$ is the speed of light, and $\vec{r}$ represents the distance
(equiv. time delay) between the quantum dot and the beamsplitter.
Upon substitution, and omitting the irrelevant time-independent factors,
we obtain:
\begin{eqnarray}
    g_{||}^{(2)}(t,\tau) &=& \frac{1}{2} 
    \langle \pi^+(t)  \pi^+(t+\delta\tau_p+\tau)
    \pi(t+\delta\tau_p+\tau) \pi(t) \rangle \nonumber\\
    &-& \frac{1}{2}  \langle \pi^+(t) \pi^+(t+\delta\tau_p+\tau)\pi(t+\tau) 
    \pi(t+\delta\tau_p) \rangle.    
    \label{visibility_raw}
\end{eqnarray}
We note that the two photons are emitted by the same quantum dot,
and therefore we only use one set of quantum dot raising and lowering
operators, although expressed at different times.
The two photons are, in general, not independent, and the
ordering of operators matters.
We have also accounted for the fact that one of the channels in the setup
is guided by the delay line, adding the time shift $\delta\tau_p$.
In this manner we have translated the photonic correlation function
into a formula expressed by the dynamical properties of the quantum dot.
To develop the cross-correlation function further, we therefore have to
consider the dynamics of the charged exciton.

The Hamiltonian of the charged exciton confined in the quantum dot,
in the language of the dipole operators, is simply
\begin{equation}
    \hat{H} = E_{1} \pi^+\pi + E_0 \pi\pi^+,
\end{equation}
where $E_{1}$ is the energy of the charged exciton, and $E_0$ is the
energy of the final-state electron.
The system is not driven coherently; the radiative decay process
is also incoherent, and therefore is not accounted for in
the Hamiltonian.
Henceforth we operate in the interaction picture defined with
the above Hamiltonian.

The dynamics of our two-level system of charged exciton is described
by the master equation
\begin{equation}
    \frac{d}{dt}\hat{\varrho} = R[\pi^+\hat{\varrho}, \pi]
    +\sum_{n=1}^2 \left(
        c_n \hat{\varrho} c_n^+ - \frac{1}{2} \left\{  
        c_n^+ c_n , \hat{\varrho} \right\} \right),
\end{equation}
where $\hat{\varrho}$ is the density matrix, the square (curly) brackets 
denote the commutator (anticommutator), 
the collapse operators $c_1 = \frac{1}{\sqrt{T_1}}\pi$ and
$c_2 = \frac{1}{\sqrt{2\tau_d}}\sigma_z$, and $\sigma_z$ is the 
$z$ Pauli matrix.
The master equation is parametrized by the incoherent
reexcitation rate $R$ (perhaps time-dependent),
the radiative lifetime $T_1$ and the pure dephasing time $\tau_d$.
Let us unfold the master equation into the Bloch equations for
the matrix elements of the elements of the density matrix:
\begin{eqnarray}
    \frac{d}{dt} \varrho_{11} &=& R \varrho_{22} - \frac{1}{T_1} \varrho_{11},
    \label{bloch1} \\
    \frac{d}{dt} \varrho_{22} &=& -R \varrho_{22} + \frac{1}{T_1} \varrho_{11},
    \label{bloch2}\\
    \frac{d}{dt} \varrho_{12} &=& - \left( \frac{1}{2T_1}
    +\frac{1}{\tau_d}\right) \varrho_{12}, \label{bloch3}
\end{eqnarray}
and the equation for $\varrho_{21}$ is a complex conjugate of that for 
$\varrho_{12}$.
We identify $\varrho_{11}(t)=p_1(t)$, the probability of the dot being
in the "filled" state (charged exciton confined), 
and $\varrho_{22}=p_0(t)$, the probability of the dot being "empty"
(holding just the final-state electron).
The time evolution of the coherence $\varrho_{12}$ is parametrized
by the coherence time $\tau_c$ (also labeled as $T_2$) defined in 
the usual manner:
\begin{equation}
    \frac{1}{\tau_c} = \frac{1}{2T_1} +\frac{1}{\tau_d}.
\end{equation}
We note that in the timing of the Bloch equations defined above, our approach 
differs somewhat from that presented in Ref.~\onlinecite{Kiraz_PRA2004}.
The equations (\ref{bloch1}), (\ref{bloch2}) are coupled and have to be
solved as a set; we have already shown these solutions for the CW and
pulsed excitation cases in the previous Sections.
The equation (\ref{bloch3}) for the coherence will now be used to
develop the interference term of the HOM two-photon visibility.

We now return to the HOM visibility, Eq. (\ref{visibility_raw})
and focus on the first term.
It expresses the probability of emission of the photon
at time $t+\delta\tau_p+\tau$ on condition that a photon has been emitted 
at time $t$.
We track the pair of dipole operators dependent on the time $\tau$, 
related to the time in which the two photons interfere on the beamsplitter.
We use the quantum regression theorem~\cite{Loudon1983,Kiraz_PRA2004}.
Since the expectation value of the product $\pi^+(t+t')\pi(t+t')$
is the probability $p_1(t+t')=\varrho_{11}(t+t')$ 
(we denote $t' = \delta\tau_p + \tau$), we need to solve
the Bloch equations (\ref{bloch1}) and (\ref{bloch2}) with the initial
condition
\begin{equation}
    \langle \pi^+(t)  \pi^+(t+t')
    \pi(t+t') \pi(t) \rangle |_{t'=0} =
    \langle \pi^+(t)  \pi^+(t) \pi(t) \pi(t) \rangle.
\end{equation}
In the case of independent single-photon sources (e.g., two independent 
quantum dots) the above value can be nonzero.
However, we implement our HOM experiment with one single-photon source,
for which the above probability of emitting two photons at the same time $t$
is zero.
As a result of this recipe we simply obtain the instantaneous
second-order correlation function
$g^{(2)}(t,\tau)$ of the charged exciton in the quantum dot
(up to normalization).
In the previous Sections we have derived the analytical formulas
for $g^{(2)}(t,\tau)$ both in the CW and pulsed excitation.

Next, we consider the interference term
$\langle \pi^+(t) \pi^+(t+\delta\tau_p+\tau)\pi(t+\tau) 
    \pi(t+\delta\tau_p) \rangle$.
Here we see four dipole operators, each at a different time, however
only two operators depend on the crucial time delay $\tau$.
Further, since each of them is taken at a different time (with the 
time difference $\delta\tau_p$), we have to treat these operators
separately.
The expectation value of the operator $\pi^+$ ($\pi$) is related
to the coherence $\varrho_{21}$ ($\varrho_{12}$) and therefore
the equation of motion for the interference term is
\begin{equation}
    \frac{d}{d\tau} \langle \pi^+(t) \pi^+(t+\delta\tau_p+\tau)\pi(t+\tau) 
    \pi(t+\delta\tau_p) \rangle 
    = -\frac{2}{\tau_c} 
    \langle \pi^+(t) \pi^+(t+\delta\tau_p+\tau)\pi(t+\tau) 
    \pi(t+\delta\tau_p) \rangle.
\end{equation}
The initial condition for the interference term is evidently (up to 
normalization factor)
\begin{eqnarray}
    \langle \pi^+(t) \pi^+(t+\delta\tau_p+\tau)\pi(t+\tau) 
    \pi(t+\delta\tau_p) \rangle|_{\tau=0}
    &=& \langle \pi^+(t) \pi^+(t+\delta\tau_p)\pi(t) 
    \pi(t+\delta\tau_p) \rangle \nonumber\\
    &=& g^{(2)}(t+\delta\tau_p).
\end{eqnarray}
After elementary calculus we find
\begin{equation}
    \langle \pi^+(t) \pi^+(t+\delta\tau_p+\tau)\pi(t+\tau) 
    \pi(t+\delta\tau_p) \rangle = g^{(2)}(t+\tau_p,\tau=0) 
    \exp\left( - \frac{2}{\tau_c} \tau  \right).
\end{equation}

We have arrived at the final expression for the 
instantaneous interference visibility:
\begin{equation}
    g_{||}^{(2)}(t,\tau) = g^{(2)}(t,\delta\tau_p+\tau)
    - g^{(2)}(t,\delta\tau_p) \exp\left( - \frac{2}{\tau_c} \tau  \right).
    \label{visibility_final}
\end{equation}
We note that the instantaneous second-order correlation 
function $g^{(2)}$ appears in both terms, although it is taken at a 
different time.
This allows us to normalize both terms with an identical factor.
Also, at the delay $\tau=0$, $g_{||}^{(2)}(t,\delta\tau_p)=0$ identically, 
which agrees with the previously published results in the electromagnetic domain~\cite{Kambs_NJP2018,Legero_AAMOP2003,Legero_APB2003,Bylander_EPJD2003}.
The closed formulas for $g_{||}^{(2)}(t,\tau)$ and its ensemble average
depend on the excitation conditions and we will summarize them below.

\subsection{Two-photon interference visibility under CW excitation}

Let us illustrate the use of the formula (\ref{visibility_final})
on the example of a system with the CW excitation, beginning with
a system of two independent single-photon sources with identical
coherence properties~\cite{Kiraz_PRA2004}.
In this case, the (unnormalized) conditional probability of photon 
emission is nonzero and independent of time, i.e., 
$g^{(2)}(t,\tau)=p_1 p_2$,
where $p_1$ is the probability of emission from the first source, and
$p_2$ is the probability of emission from the second source.
In such case, the normalized HOM visibility attains the well-known form
\begin{equation}
    g_{||}^{(2)}(t,\tau) = g_{||}^{(2)}(\tau) = 
    \frac{1}{2} \left[ 1 - \exp\left( -\frac{2}{\tau_c} \tau \right)\right].
\end{equation}

Let us now consider the system described in the main text,
consisting of one single-photon source, with one of the photons
guided through the mode imposing the delay $\delta\tau_p$.
We assume further that the system has reached its steady
state, i.e., the second-order correlation function
does not depend on the time $t$ of emission of the first photon of 
the pair, but depends only on the overall delay $\delta\tau_p+\tau$.
If the delay $\delta\tau_p$ is sufficiently long, then we may 
approximate
\begin{equation}
    g^{(2)}(\delta\tau_p) \approx g^{(2)}(\delta\tau_p+\tau)
\end{equation}
and the HOM visibility becomes 
\begin{equation}
    g_{||}^{(2)}(t,\tau) = g_{||}^{(2)}(\tau) = 
    \frac{1}{2} g^{(2)}(\delta\tau_p+\tau)
    \left[ 1 - \exp\left( -\frac{2}{\tau_c} \tau \right)\right].
\end{equation}
The HOM visibility has a factor originating from the
single-photon nature of the photon source.
However, owing to the two emission events being separated by a
long delay (at least $\delta\tau_p$), the photons are approximately
uncorrelated, and the system has coherence characteristics as if 
it involved two independent photon sources.
This assumption is made in the main text, however, it is adjusted
to account for the subtleties of the two-beamsplitter setup.

Finally, if the delay $\delta\tau_p$ is not long (comparable
to the coherence time $\tau_c$),
the complexities of conditional two-photon emission
have to be taken into account, and we have
\begin{equation}
    g_{||}^{(2)}(t,\tau) = g_{||}^{(2)}(\tau) = 
    \frac{1}{2} \left[ g^{(2)}(\delta\tau_p+\tau)
      - g^{(2)}(\delta\tau_p)\exp\left( -\frac{2}{\tau_c} \tau 
      \right)\right].
\end{equation}
In the two latter cases we utilize the formula (\ref{g2_CW})
for $g^{(2)}(\tau)$ obtained in Section~\ref{secCW}.

\subsection{Two-photon interference visibility under pulsed excitation}

Under pulsed excitation, the reexcitation rate $R$ in Bloch equations
(\ref{bloch1}), (\ref{bloch2}) is time-dependent and periodic.
In our system, the delay $\delta\tau_p$ between the two photons is chosen to
correspond precisely to the time interval $T$ between pulses.
As a result, we first consider the instantaneous HOM visibility
expressed by Eq. (\ref{visibility_final}), with the instantaneous
second-order correlation function $g^{(2)}(t_1,\tau)$ given by
Eq. (\ref{pulsed_inst_g2}).
To enable comparison with the experimental data, we perform the
ensemble average:
\begin{equation}
    g_{||}^{(2)}(\tau) = \frac{1}{T} \int_{-\infty}^{\infty}
    g_{||}^{(2)}(t_1,\tau) dt_1.
\end{equation}
We obtain
\begin{equation}
    g_{||}^{(2)}(\tau) = 
    \frac{1}{2} \left[ g^{(2)}(T+\tau)
      - g^{(2)}(T)\exp\left( -\frac{2}{\tau_c} \tau 
      \right)\right],
\end{equation}
i.e., the formula equivalent to that used in CW excitation
with short delays $\delta\tau_d$.
However, here the second-order correlation function 
$g^{(2)}(\tau)$ attains the far more complex, time-periodic 
form (\ref{pulsed_average_g2}) developed in Sec.~\ref{secPulsed}.
This formula, modified to account for the subtleties of the 
two-beamsplitter setup, is used in the main text.

We conclude this presentation by noting that the case of pulsed
excitation of two independent single-photon sources was considered
by Kiraz~et.al.~\cite{Kiraz_PRA2004}.
The model described therein required a much simpler calculation
owing to the possibility of separating the operator time evolution
into two pairs (for each emitter separately).
Nevertheless, the results obtained are similar to those reported here.
This is due to the fact that in our system, each pulse resets
the dynamics of the quantum dot.
As a result, the HOM interference of two photons emitted in our quantum dot
after different pulses is formally equivalent to that of
two photons produced by independent, but identical emitters.
Our system is nontrivially different only if both photons can be
emitted after the same pulse (although not at the same time).
This case is, however, beyond the scope of this work.

\section{Additional Figures}

In the following we provide additional figures supporting the experiments in the main text.

\begin{figure}[hbt!]
    \centering
    \includegraphics[width=15cm,clip=true]{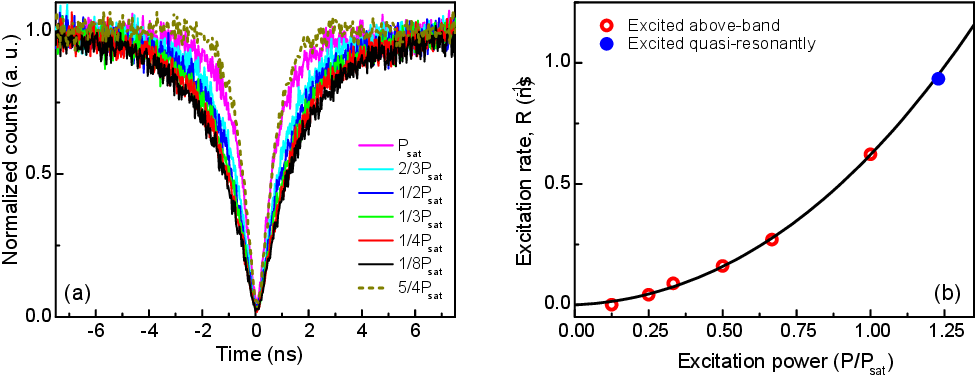}
    \caption{(a) Second-order correlation, $g^{(2)}(\tau)$, as a function of CW excitation power. Solid (dashed) lines correspond to measurements made using above-band (quasi-resonant) excitation. (b) Extracted excitation rate, $R$, using $g^{(2)}(\tau)=1-e^{-(1/T_1+R)|\tau|}$ where $T_1$ is the radiative lifetime. The formula is derived in Section~\ref{theory}.}
    \label{fig:S1}
\end{figure}

\begin{figure}
    \centering
    \includegraphics[width=12cm,clip=true]{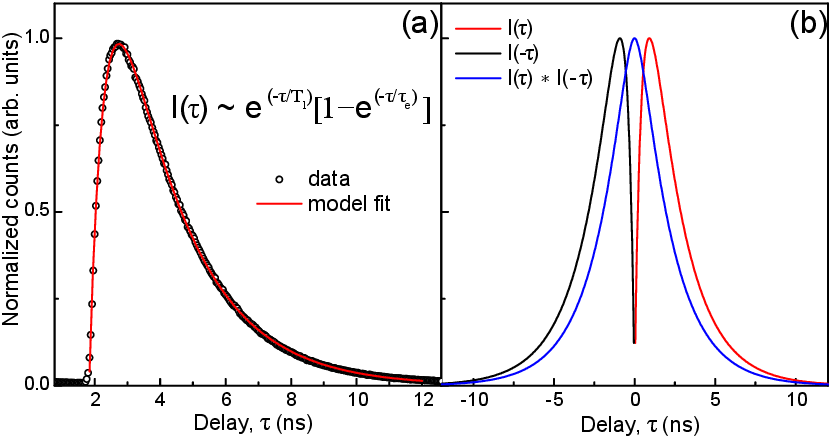}
    \caption{(a) Time-resolved photoluminescence of the $X^{1-}$ emission excited quasi-resonantly at $P=P_\mathrm{sat}$ (open symbols). The radiative lifetime, $T_1$, and the state preparation time, $\tau_e$, is extracted from the model fit of the PL decay (red curve). (b) Temporal profile of a $g^{(2)}(\tau)$ side peak constructed from the self-convolution of the model fit to the radiative decay. This approach is equivalent to the ensemble-averaged calculation of conditional probabilities for the two photons described in Section~\ref{theory}.}
    \label{fig:S2}
\end{figure}

\clearpage

\bibliography{whiskers}